\documentclass[twocolumn,showpacs,superscriptaddress]{revtex4}
\usepackage{graphicx}
\usepackage{amssymb}
\usepackage{amsfonts}
\usepackage{amsmath}
\usepackage{mathrsfs}

\begin{document}

\title{Entanglement distribution over the subsystems and its invariance}
\author{Qing-Jun Tong}
\affiliation{Center for Interdisciplinary Studies $\&$ Key Laboratory for Magnetism and Magnetic Materials of the MoE, Lanzhou
University, Lanzhou 730000, China}
\affiliation{Centre for Quantum Technologies, National University of Singapore, 3 Science Drive 2, Singapore 117543}
\author{Jun-Hong An}
\email{anjhong@lzu.edu.cn}
\affiliation{Center for Interdisciplinary
Studies $\&$ Key Laboratory for Magnetism and Magnetic Materials of the MoE, Lanzhou University, Lanzhou 730000, China}
\affiliation{Centre for Quantum Technologies, National University of Singapore, 3 Science Drive 2, Singapore 117543}
\author{Hong-Gang Luo}
\affiliation{Center for Interdisciplinary Studies $\&$ Key Laboratory for Magnetism and Magnetic Materials of the MoE, Lanzhou University, Lanzhou 730000, China}
\affiliation{Beijing Computational Science Research Center, Beijing, 100084, China}
\author{C. H. Oh}\email{phyohch@nus.edu.sg}\affiliation{Centre for Quantum Technologies, National University of Singapore, 3 Science Drive 2, Singapore 117543}

\begin{abstract}

We study the entanglement dynamics of two qubits, each of which is
embedded into its local amplitude-damping reservoir, and the entanglement distribution
among all the bipartite subsystems including qubit-qubit,
qubit-reservoir, and reservoir-reservoir. It is found that the
entanglement can be stably distributed among all components, which
is much different to the result obtained under the Born-Markovian
approximation by C. E. L\'{o}pez {\it et al.} [Phys. Rev.
Lett. \textbf{101}, 080503 (2008)], and particularly it also
satisfies an identity. Our unified
treatment includes the previous results as special cases. The result
may give help to understand the physical nature of entanglement
under decoherence.\end{abstract}


\maketitle

\section{Introduction}
Entanglement in quantum multipartite systems is a unique property in
quantum world. It plays an important role in quantum information
processing \cite{Nielsen00,Horodecki09}. Therefore, the study of its essential
features and dynamical behavior under the ubiquitous decoherence of
relevant quantum system has attracted much attention in recent years
\cite{Zyczkowski01,Yu04,Eberly,Almeida07,Laurat07,Bellomo07,Maniscalco08,Yeo10,Xu10,Xu102,Bellomo08,Wang08,Li09,An09,An092,Tong09}.
For example, it was found that the entanglement of qubits under the
Markovian decoherence can be terminated in a finite time despite the
coherence of single qubit losing in an asymptotical manner
\cite{Zyczkowski01}. The phenomenon called as entanglement sudden
death (ESD) \cite{Yu04,Eberly} has been observed experimentally
\cite{Almeida07,Laurat07}. This is detrimental to the practical realization of quantum information processing using entanglement. Surprisingly, some further studies
indicated that ESD is not always the eventual fate of the qubit
entanglement. It was found that the entanglement can revive again
after some time of ESD \cite{Bellomo07,Maniscalco08,Yeo10}, which has been
observed in optical system \cite{Xu10,Xu102}. It has been proven that this revived entanglement plays a constructive role in quantum information protocols \cite{Yeo10}. Even in some occasions, ESD
does not happen at all, instead finite residual entanglement can be
preserved in the long time limit \cite{Bellomo08,Wang08,Li09,An09}.
This can be due to the structured environment and physically it
results from the formation of a bound state between the qubit and
its amplitude damping reservoir \cite{An092,Tong09}. These results show rich dynamical
behaviors of the entanglement and its characters actually have not
been clearly identified.

Recently, L\'{o}pez \textit{et al.} asked a question about where the
lost entanglement of the qubits goes \cite{Lopez08}.
Interestingly, they found that the lost entanglement of the qubits is exclusively
transferred to the reservoirs under the Markovian amplitude-damping decoherence dynamics and ESD
of the qubits is always accompanied with the entanglement sudden
birth (ESB) of the reservoirs.
A similar situation happens for the spin entanglement
when the spin degree of freedom for one of the two particles
interacts with its momentum degree of freedom \cite{Lamata06}. All
these results mean that the entanglement does not go away, it is
still there but just changes the location. This is reminiscent of
the work of Yonac \textit{et al.} \cite{Yonac}, in which the
entanglement dynamics has been studied in a double Jaynes-Cummings (J-C) model. They
found that the entanglement is transferred periodically among all
the bipartite partitions of the whole system but an identity (see
below) has been satisfied at any time. This may be not surprising
since the double J-C model has no decoherence and any initial
information can be preserved in the time evolution. However, it
would be surprising if the identity is still valid in the presence
of the decoherence, in which a non-equilibrium relaxation process is
involved. In this paper, we show that it is indeed true for such a
system consisted of two qubits locally interacting with two amplitude-damping
reservoirs. It is noted that although the infinite degrees of
freedom of the reserviors introduce the irreversibility to the
subsystems, this result is still reasonable based on the fact that
the global system evolves in a unitary way. Furthermore, we find that the
distribution of the entanglement among the bipartite subsystems is
dependent of the explicit property of the reservoir and its coupling
to the qubit. The rich dynamical behaviors obtained previously in
the literature can be regarded as the special cases of our present
result or Markovian approximation. Particularly, we find that,
instead of entirely transferred to the reservoirs, the
entanglement can be stably distributed among all the bipartite
subsystems if the qubit and its reservoir can form a bound state and
the non-Markovian effect is important, and the
ESD of the qubits is not always accompanied with the occurrence of ESB of
reservoirs. Irrespective of how the entanglement distributes, it is
found that the identity about the entanglement in the whole
system can be satisfied at any time, which reveals the profound
physics of the entanglement dynamics under decoherence.

This paper is organized as follows. In Sec. \ref{model}, the model of two independent qubits in two local reservoirs is given. And the dynamical entanglement invariance is obtained based on the exact solution of the non-Markovian decoherence dynamics of the qubit system.  In Sec. \ref{edd}, the entanglement distribution over the subsystems when the reservoirs are PBG mediums is studied explicitly. A stable entanglement-distribution configuration is found in the non-Markovian dynamics. Finally, a brief discussion and summary are given in Sec. \ref{sum}.

\section{The model and the dynamical entanglement invariance}\label{model}
We consider two qubits interacting with two uncorrelated vacuum
reservoirs. Due to the dynamical independence between the two local
subsystems, we can firstly solve the single subsystem, then apply the result obtained to the double-qubit case. The Hamiltonian of each local subsystem is \cite{Scully97}
\begin{equation}
H=\omega _{0}\sigma _{+}\sigma _{-}+\sum_{k}\omega _{k}a_{k}^{\dag
}a_{k}+\sum_{k}(g_{k}\sigma _{+}a_{k}+h.c.),  \label{t1}
\end{equation}
where $\sigma _{\pm }$ and $\omega _{0}$ are the inversion operators and
transition frequency of the qubit, $a_{k}^{\dag }$ and $a_{k}$ are the
creation and annihilation operators of the $k$-th mode with frequency $
\omega _{k}$ of the radiation field. The coupling strength between the qubit
and the reservoir is denoted by $g_{k}=\omega_0\mathbf{\hat{e}} _{k}\cdot
\mathbf{d}/\sqrt{2\varepsilon _{0}\omega _{k}V}$, where $\mathbf{\hat{e}}
_{k} $ and $V$ are the unit polarization vector and the normalization volume
of the radiation field, $\mathbf{d}$ is the dipole moment of the qubit, and $
\varepsilon _{0}$ is the free space permittivity.

For such a system, if the qubit is in its ground state $|-\rangle$
and the reservoir is in vacuum state at the initial time, then
the system does not evolve to other states. When the qubit is in its
excited state $|+\rangle$, the system evolves as
\begin{equation}
\left\vert \phi (t)\right\rangle =b(t)\left\vert +,\{0\}_{k}\right\rangle
+\sum_{k}b_{k}(t)\left\vert -,\{1\}_{k}\right\rangle.  \label{t9}
\end{equation}
Here $\left\vert-, \{1\}_{k}\right\rangle $ denotes that the qubit
jumps to its ground state and one photon is excited in the $k$-th
mode of the reservoir. $b(t)$ satisfies an integro-differential
equation
\begin{equation}
\dot{b}(t)+i\omega _{0}b(t)+\int_{0}^{t}b(\tau )f(t-\tau )d\tau=0 ,
\label{t7}
\end{equation}
where the kernel function $f(t-\tau)=\int_{0}^{\infty }d\omega
J(\omega )e^{-i\omega (t-\tau)}$ is dependent of the spectral
density $J(\omega )=\sum_{k}\left\vert g_{k}\right\vert ^{2}\delta
(\omega -\omega _{k})$. Introducing the normalized collective state
of the reservoir with one excitation as $\left\vert
\mathbf{\tilde{1}} \right\rangle
_{r}=\frac{1}{\tilde{b}(t)}\sum_{k}b_{k}(t)\left\vert
\{1\}_{k}\right\rangle$ and with zero excitation as $\left\vert
\mathbf{ \tilde{0}}\right\rangle _{r}=\left\vert
\{0\}_{k}\right\rangle $ \cite{Lopez08}, Eq. (\ref{t9}) can be
written as $ \left\vert \phi (t)\right\rangle =b(t)\left\vert
+\right\rangle \left\vert \mathbf{\tilde{0}}\right\rangle
_{r}+\tilde{b}(t)\left\vert -\right\rangle \left\vert
\mathbf{\tilde{1}}\right\rangle _{r}$, where
$\tilde{b}(t)=\sqrt{1-\left\vert b(t)\right\vert ^{2}}$. It should
be emphasized that the introducing of normalized collective state is
not a reduction of present model to the J-C model \cite{Yonac}, as
noted in \cite{Lopez08}. The dynamics is given by Eq. (\ref{t7}),
which is difficult to obtain analytically since its non-Markovian
nature. In general the numerical integration should be used.

It is emphasized that our treatment to the dynamics of the system is exact without resorting to the widely used Born-Markovian approximation. To compare with the conventional approximate result, we may derive straightforwardly the master equation from Eq. (\ref{t9}) after tracing over the degree of freedom of the reservoir \cite{Breuer02},
\begin{eqnarray}
\dot{\rho}(t) &=&-i\Delta (t)[\sigma _{+}\sigma _{-},\rho
(t)]+\Gamma
(t)[2\sigma _{-}\rho (t)\sigma _{+}  \nonumber \\
&&-\sigma _{+}\sigma _{-}\rho (t)-\rho (t)\sigma _{+}\sigma _{-}],
\label{mstt}
\end{eqnarray}%
where the time-dependent parameters are given by
$\Delta (t)=-\text{Im}[\frac{\dot{b}(t)}{b(t)}],~\Gamma (t)=-\text{Re}[\frac{%
\dot{b}(t)}{b(t)}]$. The time-dependent parameters $\Delta (t)$ and $
\Gamma (t)$ play the roles of Lamb shifted frequency and decay rate
of the qubit, respectively. The integro-differential equation
(\ref{t7}) contains the memory effect of the reservoir registered in
the time-nonlocal kernel function and thus the dynamics of qubit
displays non-Markovian effect. If the time-nonlocal kernel function
is replaced by a time-local one, then Eq. (\ref{mstt}) recovers the
conventional master equation under Born-Markovian approximation \cite{Chen10}.

According to the above results, the time evolution of a system consisted of two such subsystems with the initial state
$\left\vert \Phi (0)\right\rangle =(\alpha \left\vert -,-\right\rangle +\beta \left\vert
+,+\right\rangle )\left\vert \mathbf{\tilde{0}}\right\rangle
_{r_{1}}\left\vert \mathbf{\tilde{0}}\right\rangle _{r_{2}}$ is given by
\begin{equation}
\left\vert \Phi (t)\right\rangle =\alpha \left\vert -,\mathbf{\tilde{0}}\right\rangle_1
\left\vert -,\mathbf{\tilde{0}}\right\rangle _{2}+\beta \left\vert \phi (t)\right\rangle
_{1}\left\vert \phi (t)\right\rangle _{2}, \label{phit}
\end{equation}
where $\alpha$ and $\beta$ are the coefficients to determine the
initial entanglement in the system. From
$\rho=|\Phi(t)\rangle\langle\Phi(t)|$, one can obtain the
time-dependent reduced density matrix of the bipartite subsystem
qubit1-qubit2 ($q_1q_2$) by tracing over the reservoir variables. It
reads
\begin{equation}
\rho _{q_1q_2}(t)=\left(
\begin{array}{cccc}
|\beta |^{2}\left\vert b(t)\right\vert ^{4} & 0 & 0 & \beta \alpha ^{\ast
}b(t)^{2} \\
0 & p & 0 & 0 \\
0 & 0 & p & 0 \\
\beta ^{\ast }\alpha b^*(t)^{2} & 0 & 0 & x
\end{array}
\right) ,
\end{equation}
where $p=\left\vert \beta b(t)\right|^2\tilde{b}(t)^2$ and
$x=1-|\beta |^{2}|b(t)|^{4}-2p$. Similarly, one can obtain the
corresponding reduced density matrices for other subsystems like
reservoir1-reservoir2 ($r_1r_2$) and qubit-reservoir ($q_1r_1$,
$q_1r_2$, $q_2r_1$, $q_2r_2$).

Using the concurrence \cite{wootters98} to quantify entanglement, we
can calculate the entanglement of each subsystem as $C_{m}=\max
\{0,Q_{m}\}$ with $Q_{m}$ for different bipartite partitions labeled
by $m$ as
\begin{eqnarray}
&&Q_{q_{1}q_{2}} = 2|\alpha\beta||b(t)|^2 - 2p,  \label{qq} \\
&&Q_{r_{1}r_{2}} = 2|\alpha\beta|\tilde{b}(t)^2 - 2p,  \label{rr} \\
&&Q_{q_{1}r_{1}} = 2|\beta|^2|b(t)|\tilde{b}(t)=Q_{q_{2}r_{2}}, \label{qr1} \\
&&Q_{q_{1}r_{2}} = 2|\alpha\beta b(t)|\tilde{b}(t) - 2p = Q_{q_{2}r_{1}}. \label{qr2}
\end{eqnarray}
One can verify that $Q_m$ in Eqs. (\ref{qq})-(\ref{qr2}) satisfy an
identity
\begin{equation}
Q_{q_{1}q_{2}}+Q_{r_{1}r_{2}}+2\left\vert \frac{\alpha}{\beta}\right\vert
Q_{q_{1}r_{1}}-2Q_{q_{1}r_{2}}=2\left\vert
\alpha \beta \right\vert,  \label{t15}
\end{equation}
where $2|\alpha\beta|$ is just the initial entanglement present in $q_1q_2$. Eq. (\ref{t15}) recovers the explicit form derived in a double J-C model \cite{Yonac} when each of the reservoirs contains only one mode, i.e. $J(\omega)=g^{2}\delta (\omega -\omega _{0})$, where the decoherence is absent and the dynamics is reversible. It is interesting that this
identity is still valid in the present model because the reservoirs
containing infinite degrees of freedom here lead to a completely
out-of-phase interaction with qubit and an irreversibility.
Furthermore, one notes that the identity is not dependent of any
detail about $b(t)$, which only determines the detailed dynamical behavior of each components in Eq. (\ref{t15}). This result manifests certain kind of invariant nature of the entanglement.

Eq. (\ref{t15}) can be intuitively understood by the global multipartite entanglement of the whole system. The global entanglement carried by the subsystem $(q_1r_1)\otimes(q_2r_2)$ can be straightforwardly calculated from Eq. (\ref{phit}) by generalized concurrence \cite{tangle} as $2|\alpha\beta|$, which, coinciding with the bipartite entanglement initially present in $q_1q_2$, just is the right hand side of Eq. (\ref{t15}). Since there is no direct interaction between $(q_1r_1)$ and $(q_2r_2)$, this global entanglement is conserved during the time evolution. From this point, our result is consistent with the one in Refs. \cite{Lopez08} and \cite{Yonac}. Another observation of Eq. (\ref{t15}) is that the different coefficients in the left hand side are essentially determined by the energy/information transfer among the local subsystems. Explicitly, in our model the total excitation number is conserved, so the energy degradation in $q_i$ with factor $b(t)$ is compensated by the energy enhancement in $r_i$ with factor $\tilde{b}(t)$. This causes that $Q_{q_1q_2}$, $Q_{r_1r_2}$, and $Q_{q_1r_2}$, in all of which the double excitation is involved, have similar form except for the different combinations of $b(t)$ and $\tilde{b}(t)$ in Eqs. ({\ref{qq}), (\ref{rr}}), and (\ref{qr2}). The dynamical consequence of the competition of the two terms in these equations causes the sudden death/birth of entanglement characterized by the presence of negative $Q$. A different case happens for $Q_{q_1r_1}$, where only single excitation is involved and no sudden death is present. With these observation, one can roughly understand why such combination in left hand side of Eq. (\ref{t15}) gives the global entanglement.

The significance of Eq. (\ref{t15}) is that it gives us a guideline to judge how the entanglement spreads out over all the bipartite partitions. It implies that entanglement is not destroyed but re-distributed among all the bipartite subsystems and this re-distribution behavior is not irregular but in certain kind of invariant manner. The similar invariant property of entanglement evolution has also been studied in Ref. \cite{Chan10}.

In the following we explicitly discuss the entanglement distribution, especially in the steady state, by taking the reservoir as a photonic band gap (PBG) medium \cite{Yablonovitch87,Lodahl} and compare it with the previous results. We will pay our attention mainly on the consequence of the non-Markovian effect on the entanglement distribution and its differences to the results in Refs. \cite{Lopez08} and \cite{Yonac}.

\section{Entanglement distribution in PBG reservoirs}\label{edd}
For the PBG medium, the dispersion relation near the upper band-edge
is given by \cite{John94}
\begin{equation}
\omega _{k}=\omega_{c}+A(k-k_{0})^{2}, \label{dispersion}
\end{equation}
where $A\approx \omega_{c}/k_{0}^{2}$, $\omega _{c}$ is the upper band-edge frequency and
$k_{0}$ is the corresponding characteristic wave vector. In this case, the kernel function has the form
\begin{equation}
f(t-\tau)=\eta\int \frac{c^3k^{2}}{\omega_k}e^{-i\omega_k (t-\tau)} dk,  \label{kn}
\end{equation}
where $\eta = \frac{\omega _{0}^{2}d^{2}}{6\pi^{2}\varepsilon
_{0}c^3 }$ is a dimensionless constant. In solving Eq. (\ref{t7})
for $b(t)$, Eq. (\ref{kn}) is evaluated numerically. Here we do not
assume that $k$ is replaced by $k_{0}$ outside of the exponential
\cite{John99,Bellomo08,Li09,Wang08}. So our result is numerically
exact. In the following we take $\omega_c$ as the unit of frequency.

\begin{figure}[tbp]
\begin{center}
\begin{tabular}{cc}
\resizebox{39mm}{!}{\includegraphics{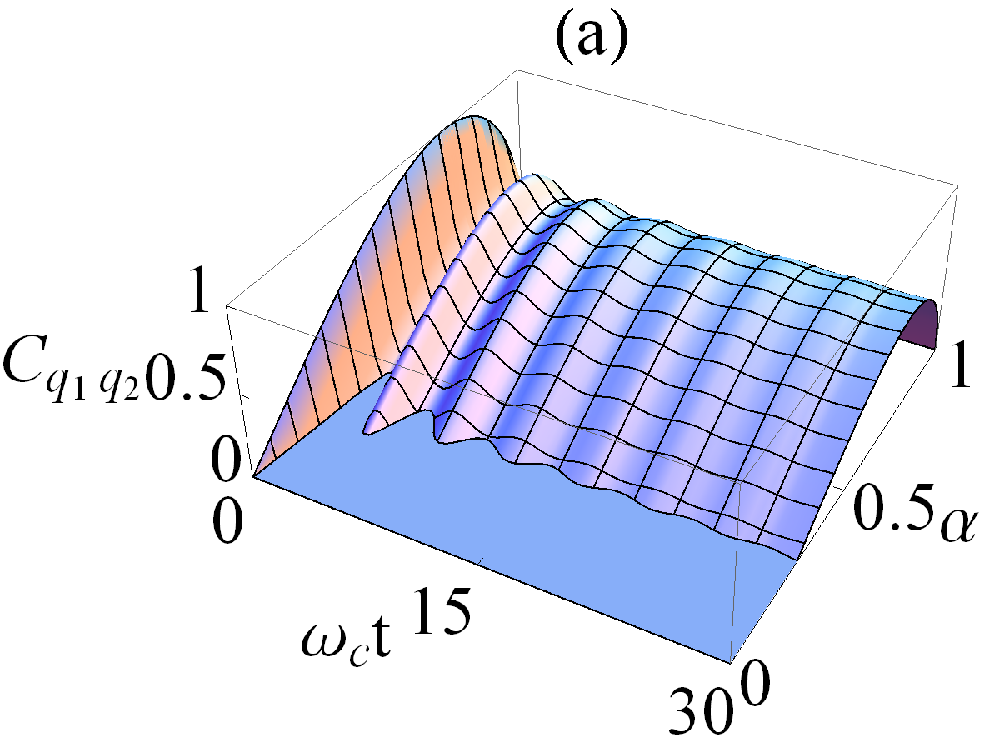}} & \resizebox{39mm}{!}{
\includegraphics{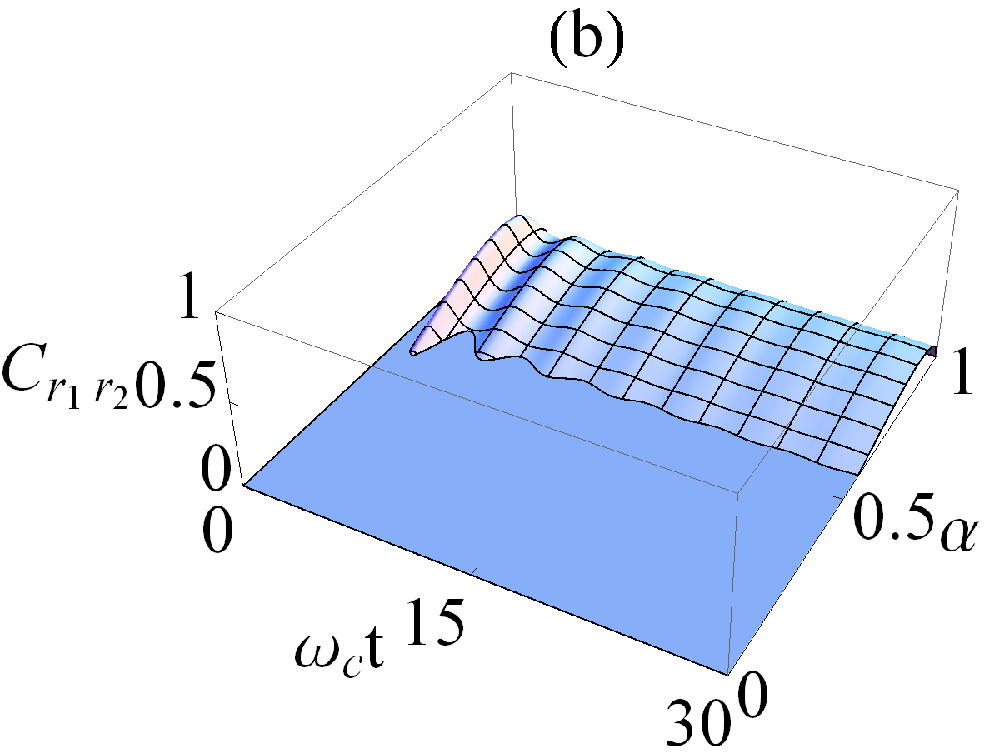}} \\
\resizebox{39mm}{!}{\includegraphics{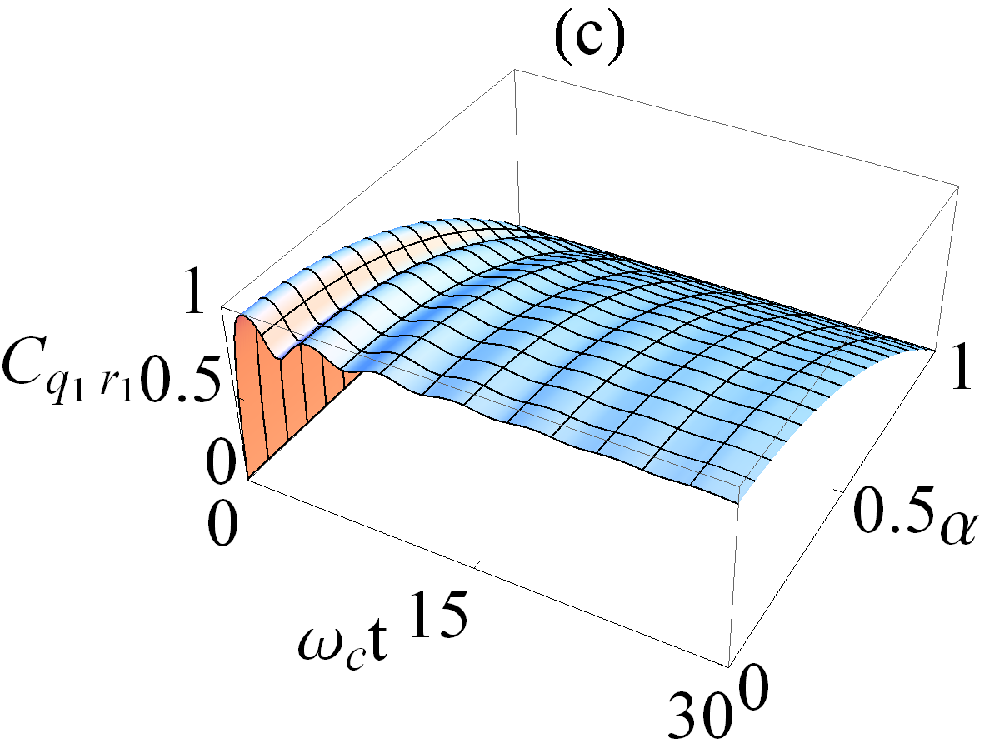}} & \resizebox{39mm}{!}{
\includegraphics{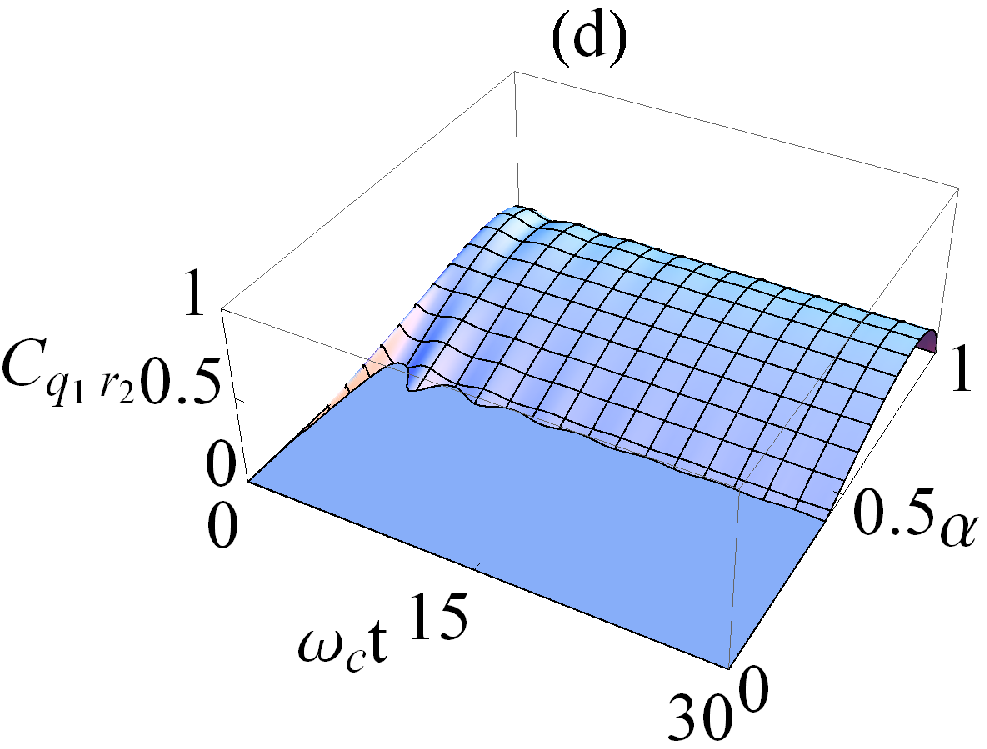}} \\
&
\end{tabular}
\end{center}
\caption{Entanglement evolutions of each bipartite partitions for
the case of $\omega_0 < \omega_c$. The parameters used are
$\omega_{0}=0.1\omega_c$ and $\eta = 0.2$.}\label{bst}
\end{figure}

\begin{figure}[tbp]
\begin{center}
\begin{tabular}{cc}
\resizebox{39mm}{!}{\includegraphics{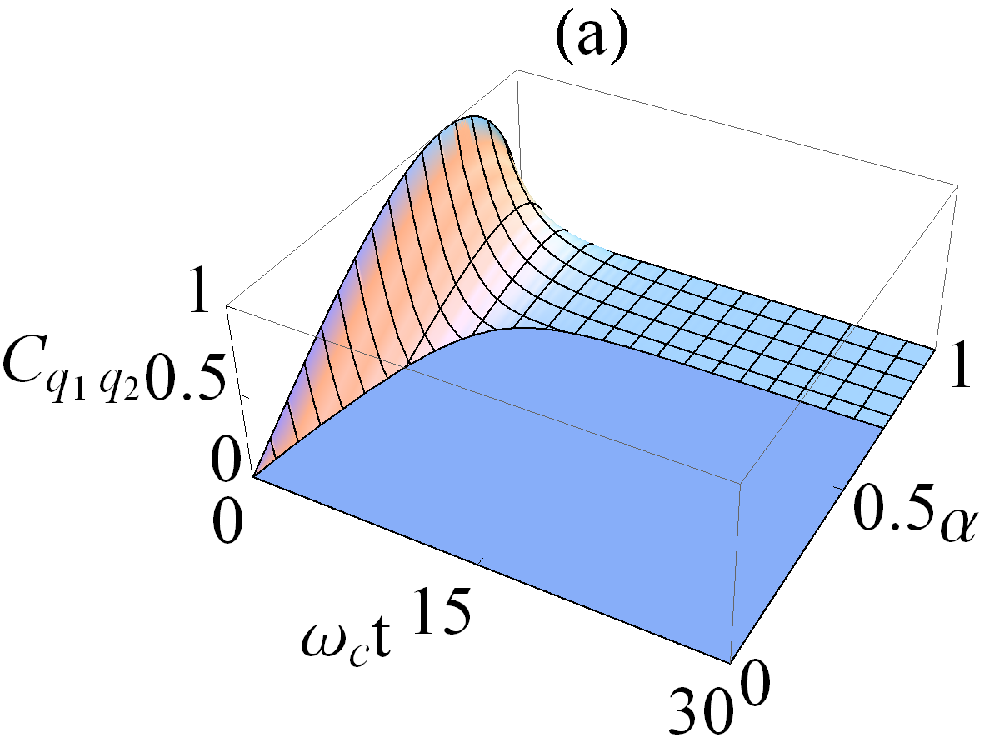}} &
\resizebox{39mm}{!}{
\includegraphics{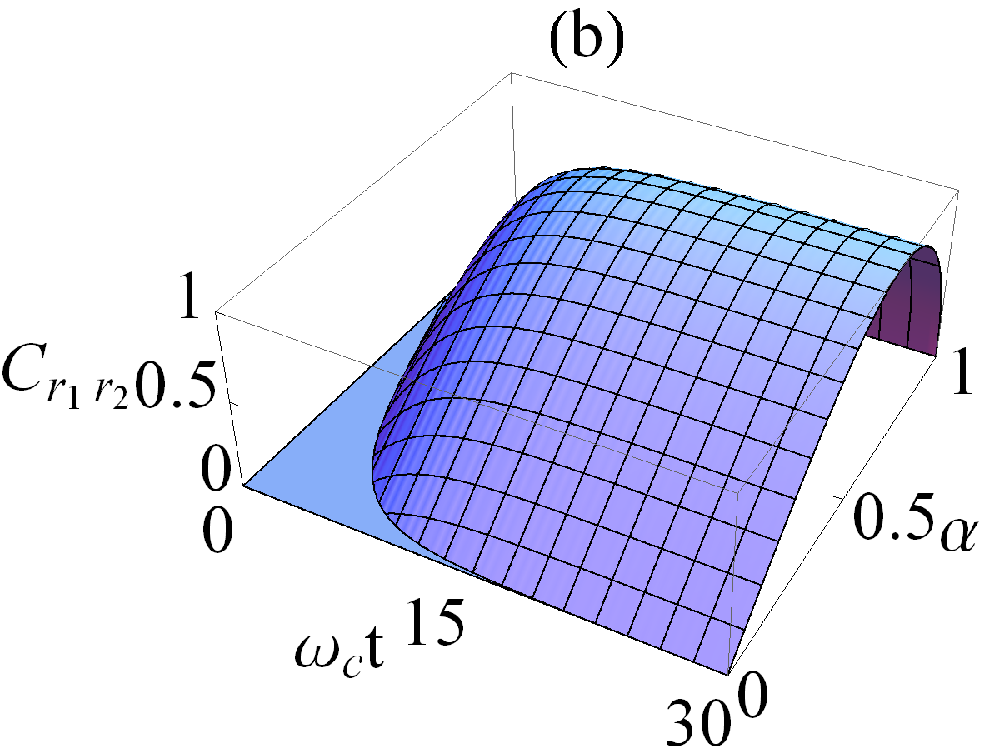}} \\
\resizebox{39mm}{!}{\includegraphics{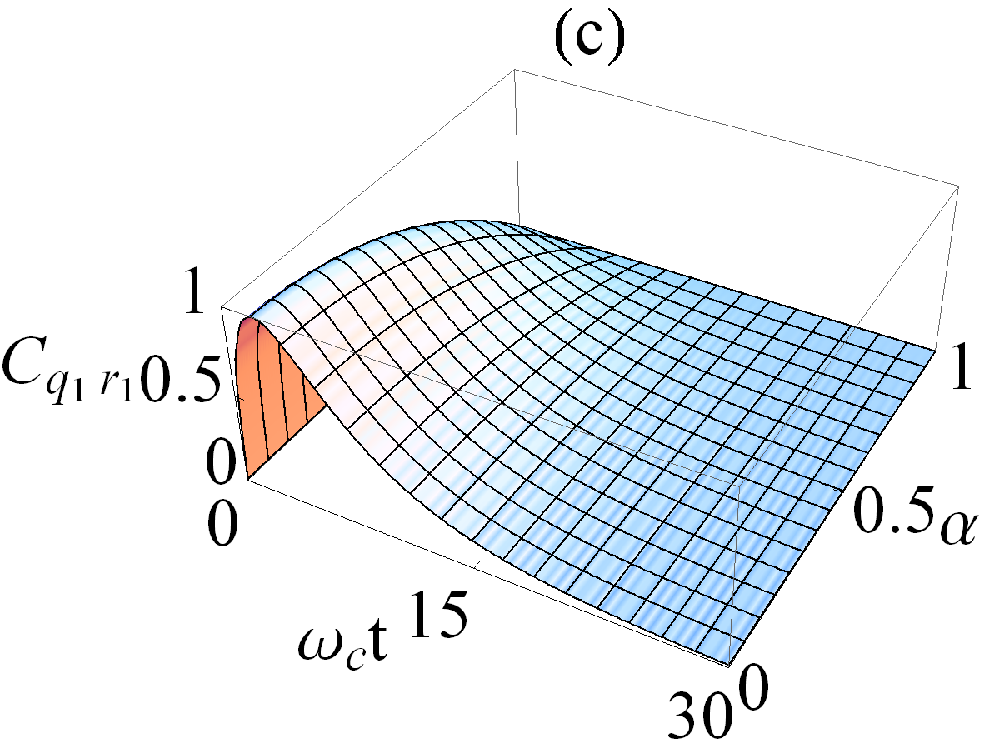}} &
\resizebox{39mm}{!}{
\includegraphics{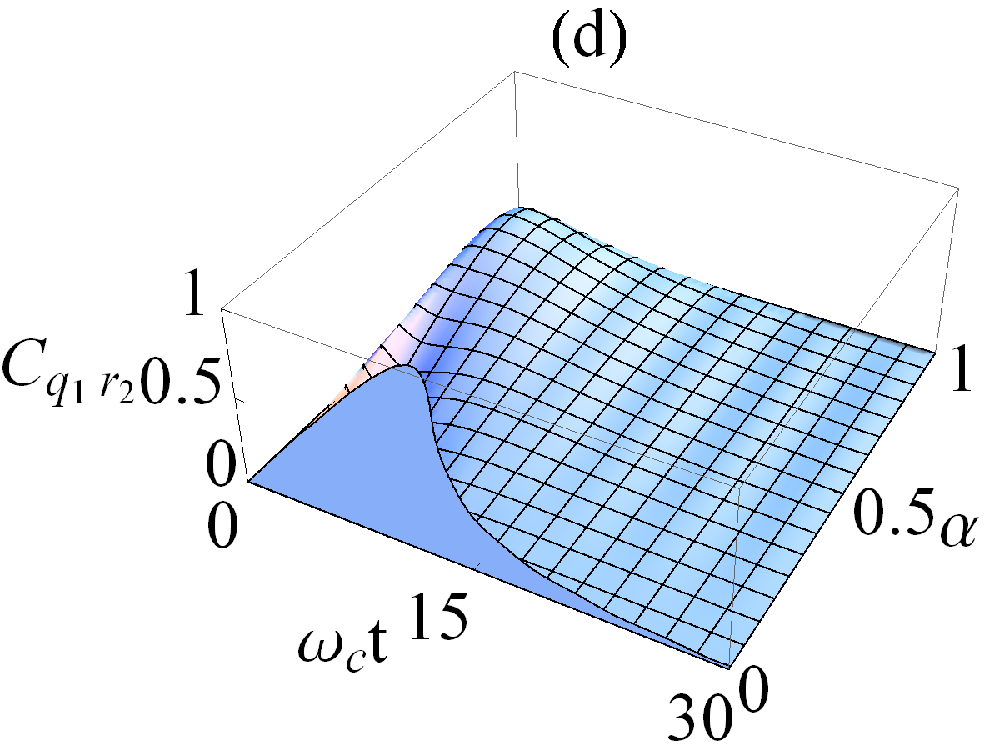}} \\
&
\end{tabular}
\end{center}
\caption{Entanglement evolutions of each bipartite partitions for
the case of $\omega_0 > \omega_c$.  The parameters used are $\omega
_{0}=10.0\omega_c$ and $\eta = 0.2$.} \label{nbst}
\end{figure}

In Figs. \ref{bst} and \ref{nbst}, we show the entanglement
evolutions of each subsystem for two typical cases of $\omega_0 <
\omega_c$ and $\omega_0
> \omega_c$, which correspond to the atomic frequency being located at the band
gap and at the upper band of the PBG medium, respectively. In the
both cases the initial entanglement in $q_1q_2$ begins to transfer
to other bipartite partitions with time but their explicit
evolutions, in particular the long time behaviors, are quite
different. In the former case, the entanglement could be distributed
stably among all possible bipartite partitions. Fig. \ref{bst}(a)
shows that after some oscillations, a sizeable entanglement of
$q_1q_2$ is preserved for the parameter regime of $0.3\lesssim
\alpha <1$. Remarkably, the entanglement in $q_ir_i(i=1,2)$ forms
quickly in the full range of $\alpha$ [Fig. \ref{bst}(c)] and
dominates the distribution. On the contrary,
only slight entanglement of $r_1r_2$ is formed in a very narrow
parameter regime $0.6\lesssim \alpha <1 $, as shown in Fig.
\ref{bst}(b). However, when $\omega_0$ is located at the upper
band of the PBG medium, the initial entanglement in $q_1q_2$ is
transferred completely to the $r_1r_2$ in the long-time limit, as
shown in Fig. \ref{nbst}. At the initial stage, $q_ir_i (i=1,2)$ and
$q_1r_2 (q_2r_1)$ are entangled transiently, but there is no stable
entanglement distribution. This result is consistent with that in
Refs. \cite{Lopez08,Zhou09,Xu09}. It is noted that the
entanglement in $q_ir_i$ comes from two parts: one
is transferred from $q_1q_2$, the other is
created by the direct interaction between $q_i$ and $r_i$. This can be seen clearly from Fig. \ref{bst}(c) when $\alpha$ is very small, the initial entanglement of $q_1q_2$ is very small, while that of $q_ir_i$ is rather large, which just results from the
interaction between $q_i$ and $r_i$. Another interesting point is a stable entanglement
can even be formed for the non-interacting bipartite
system $q_1r_2$ [see \ref{bst}(d) when $0.5\lesssim \alpha <1 $]. This entanglement transfer also results from the local interaction between $q_i$ and $r_i$ \cite{Lopez08}.

\begin{figure}[h]
\begin{center}
\includegraphics[width = 0.7 \columnwidth]{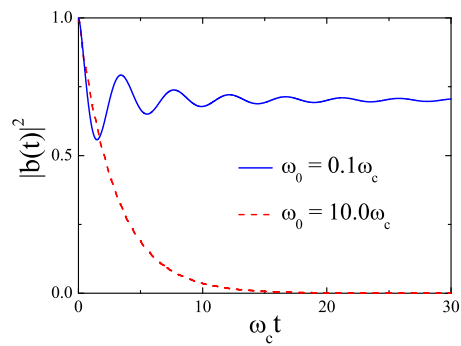}
\end{center}
\caption{Time evolution of time-dependent factor of the
excited-state population for two parameter regimes $\omega_0 =
0.1\omega_c$ (solid line) and $10.0\omega_c$ (dashed line). $\eta$
is taken as 0.2. } \label{tong0}
\end{figure}

It is not difficult to understand
these rich behaviors of entanglement distribution according to Eqs. (\ref{qq})-(\ref{qr2}) and its invariance (\ref{t15}). From these
equations, one can clearly see that the entanglement dynamics and
its distributions in the bipartite partitions are completely
determined by the time-dependent factor $|b(t)|^{2}$ of single-qubit
excited-state population. Fig. \ref{tong0} shows its time evolutions
for the corresponding parameter regimes presented above. We notice
that $|b(\infty)|^2\neq 0$ when $\omega_0$ is located at the band
gap, which means that there is some excited-state population in the
long-time limit. This phenomenon known as population trapping
\cite{Lambropoulos00} is responsible for the suppression of the
spontaneous emission of two-level system in PBG reservoir and has
been experimentally observed \cite{Yablonovitch87,Lodahl,Dreisow08}.
Such population trapping just manifests the formation of bound
states between $q_i$ and $r_i$ \cite{Tong09}, which has been
experimentally verified in \cite{Dreisow08}. Consequently, $q_i$ and
$r_i$ are so correlated in the bound states that the initial
entanglement in $q_1q_2$ cannot be fully transferred to $r_1r_2$.
The oscillation during the evolution is just the manifestation of
the strong non-Markovian effect induced by the reservoirs. On the
contrary, if $\omega_0$ is located in the upper band, then
$|b(\infty)|^2=0$ and the qubits decay completely to their ground
states. In this case the bound states between $q_i$ and $r_i$ are
absent and, according to Eq. (\ref{t15}), the initial entanglement in $q_1q_2$ is completely
transferred to $r_1r_2$, as clearly shown in Eq. (\ref{rr}).

\begin{figure}[h]
\begin{center}
\includegraphics[width = \columnwidth]{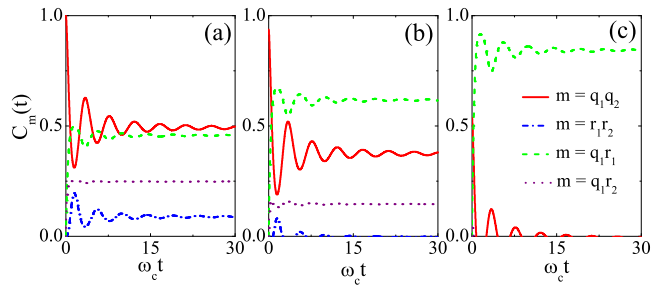}
\end{center}
\caption{(Color online) Entanglement evolution when $\alpha=1/%
\sqrt{2}$ (a), $\alpha=0.57$ (b), and $\alpha=0.28$ (c). The
parameters used here are the same as Fig. \ref{bst}.} \label{td}
\end{figure}

\begin{figure}[h]
\begin{center}
\includegraphics[width = \columnwidth]{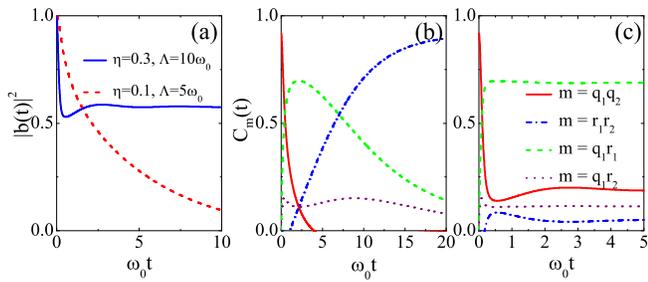}
\end{center}
\caption{(Color online) Entanglement evolution for the Ohmic
spectral density. The two sets of parameters $(\eta,\Lambda) = (0.1,
5\omega_0)$ and $(0.3, 10\omega_0)$ have been considered for
comparison. The corresponding entanglement evolutions are given in
(b) and (c), respectively. In both cases $\alpha=0.55$.} \label{osd}
\end{figure}

In addition, in Refs. \cite{Lopez08,Zhou09,Xu09} it was emphasized
that ESD of $q_1q_2$ is always accompanied with ESB of $r_1r_2$.
However, this is not always true. To clarify this, we examine the
condition to obtain ESD of the qubits and the companying ESB of the
reservoirs. From Eqs. (\ref{qq}) and (\ref{rr}) it is obvious that
the condition is $Q_{q_{1}q_{2}}(t)<0$ and $
Q_{r_{1}r_{2}}(t^{\prime })>0$ at any $t$ and $t^{\prime }$, which
means
\begin{equation}
\left\vert b(t^{\prime })\right\vert ^{2}<| \alpha | /\sqrt{1-|\alpha | ^{2}}
<1-\left\vert b(t)\right\vert ^{2}.  \label{dd}
\end{equation}
When the bound states is absent, $\left\vert b(\infty )\right\vert
^{2}=0$, the condition (\ref{dd}) can be satisfied when
$\alpha<1/\sqrt{2}$. So one can always expect ESD of the qubits and
the companying ESB of the reservoirs in the region $\left\vert
\alpha \right\vert <1/\sqrt{2}$, as shown in Fig. \ref{nbst} and
Refs. \cite{Lopez08,Zhou09,Xu09}. However, when the bound states are
available, the situation changes. In particular, when $\left\vert
b(t)\right\vert ^{2}\geq \frac{1}{2}$ in the full range of time
evolution, no region of $\alpha $ can make the condition (\ref{dd})
to be satisfied anymore. For clarification, we present three typical
behaviors of the entanglement distribution in Fig. \ref{td}. In all
these cases the bound states are available. Fig. \ref{td}(a) shows
the situation where the entanglement is stably distributed among all
of the bipartite subsystems. Fig. \ref{td}(b) indicates that the
entanglement of $ r_1r_2$ shows ESB and revival, while the
entanglement of $q_1q_2$ does not exhibit ESD. Fig. \ref{td}(c)
shows another example that while the entanglement of $q_1q_2$ has
ESD and revival \cite{Bellomo07}, the entanglement of $r_1r_2$ does
not show ESB but remains to be zero. Both Fig. \ref{td}(b,c) reveal
that ESD in $q_1q_2$ has no direct relationship with ESB in
$r_1r_2$.

\section{Discussion and summary}\label{sum}
%The above discussion is general and is not dependent of the explicit
%form of the reservoir.
The above discussion is not dependent of the explicit spectral density of the
individual reservoir. To confirm this, we consider the reservoir in
free space. The spectral density has the Ohmic form $J(\omega)=\eta
\omega \exp (-\omega/\Lambda)$, which can be obtained from the
free-space dispersion relation $\omega=ck$. One can verify that the
condition for the formation of bound states is: $\omega_0-\eta
\Lambda<0$ \cite{Tong09}. In Fig. \ref{osd}, we plot the results in
this situation. The previous results can be recovered when the bound
states are absent \cite{Lopez08}. On the contrary, when the bound
states are available, a stable entanglement is established among all
the bipartite partitions. Therefore, we argue that the stable
entanglement distribution resulted from the bound states is a
general phenomenon in open quantum system when the non-Markovian
effect is taken into account.

In summary, we have studied the entanglement distribution among all
the bipartite subsystems of two qubits embedded into two independent amplitude damping
reservoirs. It is found that the entanglement can be stably
distributed in all the bipartite subsystems, which is much different no matter to the Markovian approximate result \cite{Lopez08} or to the decoherenceless double J-C model result \cite{Yonac}, and an identity about
the entanglement in all subsystems is always satisfied. This
identity is shown to be independent of any detail of the reservoirs
and their coupling to the qubit, which affect only the explicit time
evolution behavior and the final distribution. The result is
significant to the study of the physical nature of entanglement
under decoherence. It implies an active way to protect entanglement from decoherence by modifying the properties of
the reservoir via the potential usage of the newly emerged technique, i.e. quantum reservoir engineering \cite{Myatt,Diehl}.
%The result shows the physical nature of the
%entanglement and has a significant implication for the quantum
%information processing.
\section*{Acknowledgments}
This work is supported by the Fundamental Research Funds for the
Central Universities under Grant No. lzujbky-2010-72, Gansu
Provincial NSF under Grant No. 0803RJZA095, the national NSF of China, the program for
NCET, and the CQT WBS grant No. R-710-000-008-271.

\end{document}